\documentclass{sig-alternate-2013}
\usepackage{array}
\usepackage{booktabs}
\newcolumntype{L}[1]{>{\raggedright\let\newline\\\arraybackslash\hspace{0pt}}m{#1}}
\newcolumntype{C}[1]{>{\centering\let\newline\\\arraybackslash\hspace{0pt}}m{#1}}
\newcolumntype{R}[1]{>{\raggedleft\let\newline\\\arraybackslash\hspace{0pt}}m{#1}}
\usepackage{hyperref}
\usepackage{multirow}
\usepackage{hanging}

\permission{Preprint version}
\copyrightetc{}

\begin{document}

% Copyright
% \setcopyright{acmcopyright}
%\setcopyright{acmlicensed}
%\setcopyright{rightsretained}
%\setcopyright{usgov}
%\setcopyright{usgovmixed}
%\setcopyright{cagov}
%\setcopyright{cagovmixed}

% \acmPrice{\$15.00}

%
% --- Author Metadata here ---
% \CopyrightYear{2016} % Allows default copyright year (20XX) to be over-ridden - IF NEED BE.
% \crdata{978-1-4503-4144-8/16/04. \\
% Include the http://DOI string/url }
% Allows default copyright data (0-89791-88-6/97/05) to be over-ridden - IF NEED BE.
% --- End of Author Metadata ---

% GLC: We decided not to use 'Hoaxy'. I haven't found a better
% alternative. Pheme (also called Ossa) and Fama are the names of 
% the greek and roman goddesses of rumor (see Wikipedia). But we cannot 
% use `Pheme' since it is already taken by the Pheme.eu project, and `Ossa'
% is too obscure. `Fama' could be an option but I don't really like it, and
% its main meaning is more `fame' than `rumor'.
% 
% Chengcheng, if you have something of your own in mind please let us know! 
% Otherwise I would just use what we have now.

\title{Hoaxy: A Platform for Tracking Online Misinformation}


\numberofauthors{2} 

\author{
\alignauthor
Chengcheng Shao\titlenote{Work performed as a visiting scholar at Indiana University. Email: shaoc@indiana.edu}\\
    \affaddr{School of Computer}\\
    \affaddr{National University of Defense Technology, China}\\
    %\affaddr{Changsha, Hunan 410073, China}%\\
    %\email{shaoc@indiana.edu}
%
\alignauthor
Giovanni Luca Ciampaglia$^1$, \\ Alessandro Flammini$^{1,2}$, \\ Filippo Menczer$^{1,2}$\\
      \affaddr{$^{1}$ Indiana University Network Science Institute}\\
      \affaddr{$^{2}$ School of Informatics and Computing}\\
      \affaddr{Indiana University, Bloomington, USA}\\
      %\affaddr{1001 Sr 45/46}\\
      %\affaddr{Bloomington, IN 47408}%\\
      %\email{gciampag@indiana.edu}
%
%\alignauthor
%Alessandro Flammini \and Filippo Menczer\\
%      \affaddr{Indiana University, Bloomington, USA}\\
      %\affaddr{919 E 10th St}\\
      %\affaddr{Bloomington, IN 47408}%\\
      %\email{\{aflammin,fil\}@indiana.edu}
}

\maketitle

\begin{abstract}
Massive amounts of misinformation have been observed to spread in uncontrolled fashion across social media. Examples include rumors, hoaxes, fake news, and conspiracy theories. At the same time, several journalistic organizations devote significant efforts to high-quality fact checking of online claims. The resulting information cascades contain instances of both accurate and inaccurate information, unfold over multiple time scales, and often reach audiences of considerable size. All these factors pose challenges for the study of the social dynamics of online news sharing. Here we introduce \emph{Hoaxy}, a platform for the collection, detection, and analysis of online misinformation and its related fact-checking efforts. We discuss the design of the platform and present a preliminary analysis of a sample of public tweets containing both fake news and fact checking. We find that, in the aggregate, the sharing of fact-checking content typically lags that of misinformation by 10--20 hours. Moreover, fake news are dominated by very active users, while fact checking is a more grass-roots activity. With the increasing risks connected to massive online misinformation, social news observatories have the potential to help researchers, journalists, and the general public understand the dynamics of real and fake news sharing.
\end{abstract}

%\begin{CCSXML}
%...
%\end{CCSXML}
%
%\ccsdesc[500]{Computer systems organization~Embedded systems}
%...


%
%  Use this command to print the description
%
% \printccsdesc


\section{Introduction}

The recent rise of social media has radically changed the way people consume and produce information online~\cite{Kaplan2010}. Approximately 65\% of the US adult population accesses the news through social media~\cite{Anderson, Perrin2015}, and more than a billion people worldwide are active on a daily basis on Facebook alone~\cite{Facebook2015}. 

%Social media fall within the broader category of user-generated content communities, like e.g.~the popular video sharing website YouTube~\cite{Cha2007}, or Wikipedia~\cite{Priedhorsky2007}. 
The possibility for normal consumers to produce content on social media creates new economies of attention~\cite{Ciampaglia2015a} and has changed the way companies relate to their customers~\cite{Tapscott2008}. Social media allow users to participate in the propagation of the news. For example, in Twitter, users can rebroadcast, or \emph{retweet}, any piece of content to their social circles, creating a competition among posts for our limited attention~\cite{weng2012competition}. 
This has the implication that no individual authority can dictate what kind of information is distributed on the whole network. While such platforms have brought about a more egalitarian model of information access according to some~\cite{benkler2006wealth}, the inevitable lack of oversight from expert journalists makes social media vulnerable to the unintentional spread of false or inaccurate information, or \emph{misinformation}.

Large amounts of misinformation have been indeed observed to spread
online in viral fashion, oftentimes with worrying consequences in the offline world~\cite{Carvalho2011, kata2012anti, Nyhan2013, Lauricella2013,
mocanu2014collective, Buttenheim2015}; examples include
rumors~\cite{friggeri2014rumor}, false
news~\cite{Carvalho2011,Lauricella2013}, hoaxes, and even elaborate
conspiracy theories~\cite{anagnostopoulos2014viral,DelVicario2016,galam2003modelling}.

Due to the magnitude of the phenomenon, media organizations are devoting increasing efforts to produce accurate verifications in a timely manner. For example, during Hurricane Sandy, false reports that the New York Stocks Exchange had been flooded were corrected in less than an hour~\cite{NYSE}. These \emph{fact-checking} assessments are consumed and broadcast by social media users like any other type of news content, leading to a complex interplay between `memes' that vie for the attention of users~\cite{weng2012competition}. Examples of such organizations include \url{Snopes.com}, PolitiFact, and \url{FactCheck.org}. 

Structural features of the information exchange networks underlying social media create peculiar patterns of information access. Online social networks are characterized by homophily~\cite{McPherson2001}, polarization~\cite{Conover2011}, algorithmic ranking~\cite{Bakshy2015}, and social bubbles~\cite{Nikolov2015} --- information environments with low content diversity and strong social reinforcement.

All of these factors, coupled with the fast news life cycle~\cite{Dezsoe2006,Ciampaglia2015a}, create considerable challenges for the study of the dynamics of social news sharing. To address some of these challenges, here we present \emph{Hoaxy}, an upcoming Web platform for the tracking of social news sharing. Its goal is to let researchers, journalists, and the general public monitor the production of online misinformation and its related fact checking.

As a simple proof of concept for the capabilities of this kind of systems, here we present the results of a preliminary analysis on a dataset of public tweets collected over the course of several months. We focus on two aspects: the temporal relation between the spread of misinformation and fact checking, and the differences in how users share them.
We find that, in absolute terms, misinformation is produced in much larger quantity than fact-checking content. Fact checks obviously lag misinformation, and we present evidence that there exists a characteristic lag of approximately 13 hours between the production of misinformation and that of fact checking. Finally, we show preliminary evidence that fact-checking information is spread by a broader plurality of users compared to fake news.  

\section{Related Work}

Tracking abuse of social media has been a topic of intense research in recent years. Beginning with the detection of simple instances of political abuse like \emph{astroturfing}~\cite{ratkiewicz2011detecting}, researchers noted the need for automated tools for monitoring social media streams. Several such systems have been proposed in recent years, each with a particular focus or a different approach. The Truthy system~\cite{Ratkiewicz2011}, which relies on network analysis techniques, is among the best known of such platforms. The TweetCred system~\cite{Castillo2011} focuses instead on content-based features and other kind of metadata, and distills a measure of overall information credibility. 

More recently, specific systems have been proposed to detect rumors. These include RumorLens~\cite{Resnick2014},  TwitterTrails~\cite{metaxas2015using}, and FactWatcher~\cite{Hassan2014}. The fact-checking capabilities of these systems range from completely automatic (TweetCred), to semi-automatic (TwitterTrails, RumorLens). In addition, some of them let the user explore the propagation of a rumor with an interactive dashboard (TwitterTrails, RumorLens). However, they do not monitor the social media stream automatically, but require the user to input a specific rumor to investigate. Compared with these, the objective of the Hoaxy system is to track both accurate and inaccurate information in automatic fashion. 

Automatic attempts to perform fact checking have been recently proposed for simple statements~\cite{Ciampaglia2015}, and for multimedia content~\cite{Boididou2014}. At this initial stage, because our focus is on automatic tracking of news sharing, the Hoaxy system does not perform any kind of fact checking. Instead, we focus on tracking news shares from sources whose accuracy has been determined independently. There have been investigations on the related problems of finding reliable information sources~\cite{Diakopoulos2012} and news curators~\cite{Lehmann2013}.

\section{System Architecture}
\label{sec:sys_arc}

\begin{figure}
\centering
\includegraphics[width=\columnwidth]{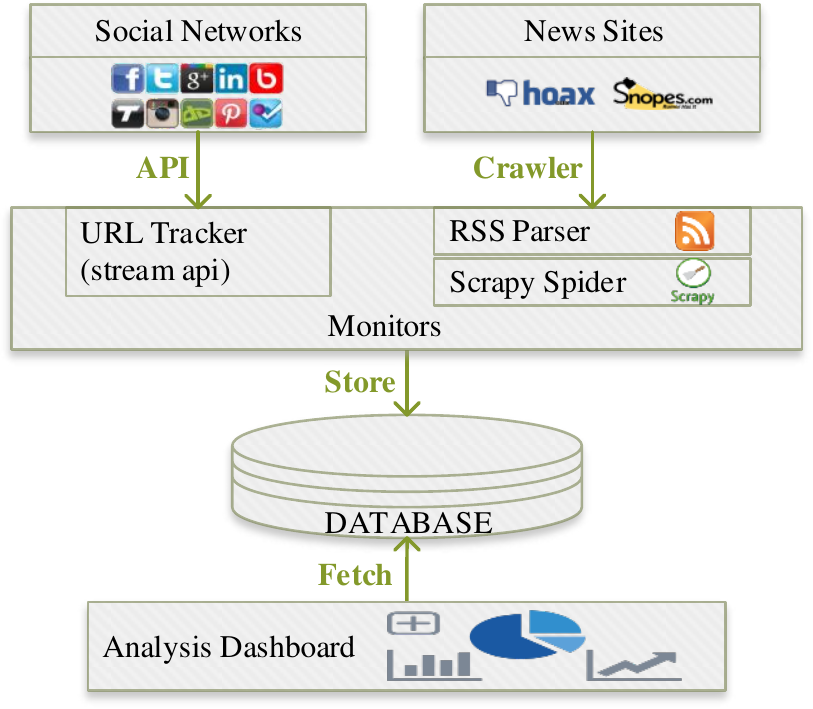}
\caption{Architecture of the Hoaxy system.}
\label{fig:sys_arc}
\end{figure}

%In this section we give a brief description of the architecture of our system. 
Our main objective is to build a uniform and extensible platform to collect and track misinformation and fact checking. Fig.~\ref{fig:sys_arc} shows the architecture of our system. Currently our efforts have been focused on the `Monitors' part of the system. We have implemented a tracker for the Twitter API, and a set of crawlers for both fake news and fact checking websites, as well as a database. 

We begin by describing the origin of our data. The system collects data from two main sources: news websites and social media. 
%(e.g., Twitter, Facebook, and Google Plus). 
From the first group we can obtain data about the origin and evolution of both fake news stories and their fact checking. From the second group we collect instances of these news stories (i.e., URLs) that are being shared online.  

To collect data from such disparate sources, we make use of a number of technologies: Web scraping, Web syndication, and, where available, APIs of social networking platforms. For example, we use the Twitter streaming API to do real-time tracking of news sharing. Because tweets are limited to 140 characters, the most common method to share a news story on Twitter is to include directly a link to its Web article. This means that we can focus only on tweets containing links to specific domains (websites), a task that is performed efficiently by the \emph{filter} endpoint of the Twitter streaming API.\footnote{\url{dev.twitter.com/streaming/reference/post/statuses/filter}}

To collect data on news stories we rely on RSS, which allows us to use a unified protocol instead of manually adapting our scraper to the multitude of Web authoring systems used on the Web. Moreover, RSS feeds contain information about updates made to news stories, which let us track the evolution of news articles. We collect data from news sites using the following two steps: when a new website is added to our list of monitored sources, we perform a `deep' crawl of its link structure using a custom Python spider written with the Scrapy framework\footnote{\url{scrapy.org}}; at this stage, we also identify the URL of the RSS feed, if available. Once all existing stories have been acquired, we perform every two hours a `light' crawl by checking its RSS feed only. To perform the `deep' crawl, we use a depth-first strategy. The `light' crawl is instead performed using a breadth-first approach. 
We store all these structured data into a database; this allows convenient retrieval for future analysis, which we plan to implement as an interactive Web dashboard. Unfortunately, URLs do not make for good, unique identifiers, since URLs with different protocol schema, query parameters, or fragments may all refers to the same page. We rely on canonical URLs where possible, and adopt a simple URL canonization technique in other cases (see below). 

\newpage
\section{Preliminary Analysis}

In this section we report results from a preliminary analysis performed on a large set of public tweets collected over the course of several months. Since we are interested in characterizing the relation between the overall social sharing activity of misinformation and fact checking, we begin our analysis by focusing on the overall aggregate volume of tweets, without breaking activity down to the level of an individual story or set of stories. We take aggregate volume as a proxy for the overall social sharing activity of news stories.

\begin{figure}[t]
\centering
\includegraphics[width=\columnwidth]{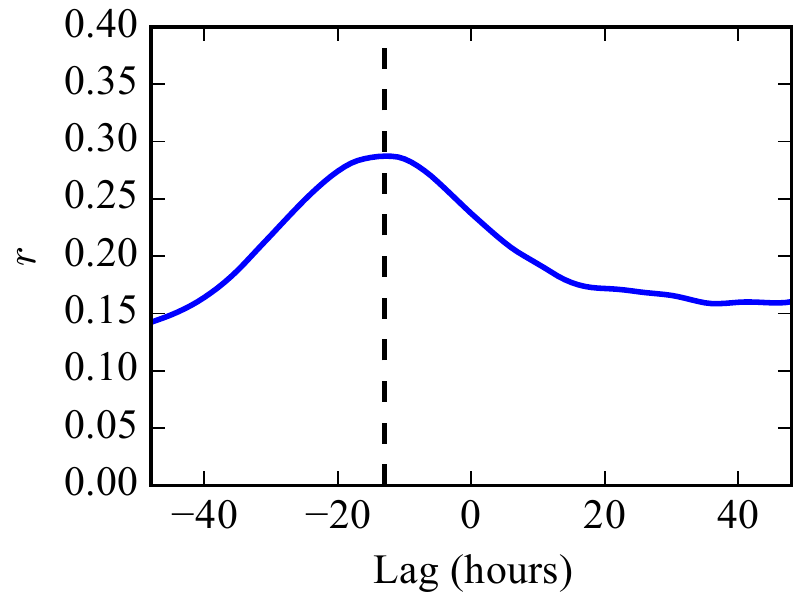}
\caption{Lagged cross correlation (Pearson's $r$) between news sharing activity of misinformation and fact-checking, with peak value at $\textnormal{lag}=-13$ hours.}
\label{fig:ccf}
\end{figure}

\subsection{Data}

We collect tweets containing URLs from two lists of Web domains: the first, fake news, covers 71 domains and was taken from a comprehensive resource on online misinformation.\footnote{\url{fakenewswatch.com}} We manually removed known satirical sources like \textit{The Onion.} The second list is composed of the six most popular fact-checking websites: \url{Snopes.com}, \url{PolitiFact.com}, \url{FactCheck.org}, \url{OpenSecrets.org}, \url{TruthOrFiction.com}, and \url{HoaxSlayer.com}. The keywords we used to collect all these tweets correspond to the domain names of these websites.

To convert the URLs to canonical form we perform the following steps: first, we transform all text into lower case; then we remove the protocol schema (e.g. `http://'); then we remove, if present, any prefix instance of the strings `www.' or `m.'; finally, we remove all URL query parameters.

\begin{table}[tb]
\centering
\caption{Summary statistics of tweet data.}
\begin{tabular}{rrrrr}
\toprule
source &   $N_\textnormal{sites}$    &  $N_\textnormal{tweets}$  &  $N_\textnormal{users}$   &  $N_\textnormal{URLs}$\\
\midrule
fake news     & 71   & 1,287,769   &  171,035   & 96,400  \\
fact checking & 6    & 154,526     &  78,624    & 11,183  \\
\bottomrule
\end{tabular}
\label{tab:data_description}
\end{table}

We collected about 3 months of filtered tweets traffic from Oct 14, 2015 to Jan 24, 2016. The summary statistics for the numbers of tweets, unique users, and unique canonical URLs (Table~\ref{tab:data_description}) illustrate the imbalance between the sets of fake news and fact-checking sites.

\begin{figure}[t]
\centering
\includegraphics[width=\columnwidth]{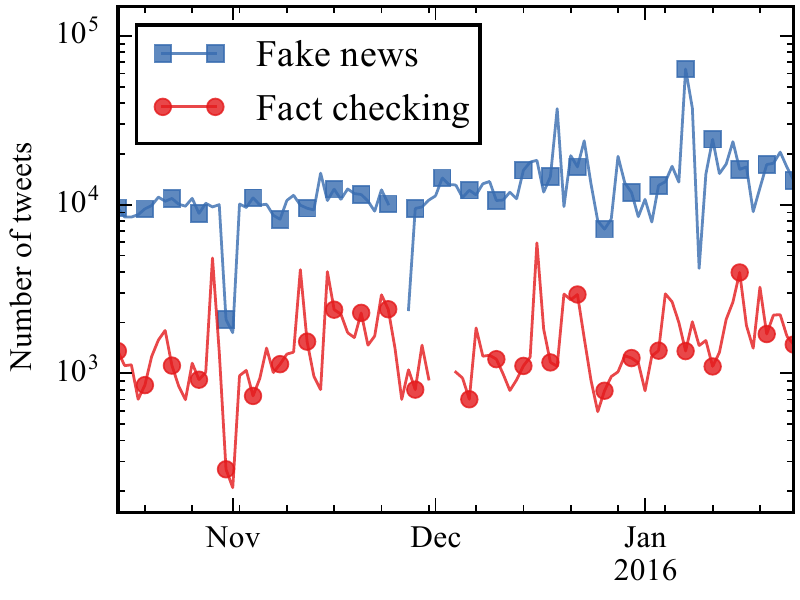}
\caption{Daily volume of tweets. The gaps correspond to two windows with missing data when our collection script crashed. 
}
\label{fig:tweets_traffic}
\end{figure}

\subsection{Tweet Volume}

Fig.~\ref{fig:tweets_traffic} plots the daily volume of tweets. As described before, we track more fake sites than fact checking ones, so the volume of fake news tweets is larger than that of fact checking ones by approximately one order of magnitude.

While both time series display significant fluctuations, the presence of aligned peaks (Nov 16) and valleys (Nov 2) suggests the presence of cross-correlated activity. To better understand this, we perform a lagged cross-correlation analysis, which measures the similarity between two time series signals as a function of the lag.

Fig.~\ref{fig:ccf} shows the results of the cross correlation analysis, with lags ranging from $-48$ hours to $+48$ hours. A higher correlation at a \emph{negative} lag indicates that the sharing of fake news \emph{precedes} that of fact checking. To eliminate circadian fluctuations, we use a simple moving average method with centered window of size equal to 24 hours. The results suggest that, in the limited number of examples at our disposal, there is a characteristic time lag between fake news and fact checking of approximately 13 hours. Because the moving average cleaning could only remove circadian fluctuations, we do not exclude the presence of correlations at larger lags (e.g., weekly). 

While this cross-correlation is suggestive of a temporal relation between misinformation and fact-checking, it is important to understand that it is based on aggregate data. 
%To shed more light on the actual news sharing processes, 
We selected a subset of URLs from both data sets to see if these correlations also hold at the level of individual events. We followed two strategies: (1) we selected a single URL from our pool of fake news stories and a matching URL from that of fact-checking stories; (2) we considered a small set of keywords and used them to perform pattern matching on the lists of URLs.

\begin{table*}[t!]
\centering
\caption{A1: An example of inaccurate news story. A2: Corresponding fact checking page. B1: News articles reporting inaccurate information about the death of actor Alan Rickman. B2: Corresponding fact-checking pages.}
\footnotesize
\begin{tabular}{L{0.16in} L{0.93\textwidth}}
\toprule
A1 & \url{www.infowars.com/white-house-gave-isis-45-minute-warning-before-bombing-oil-tankers/} \\
\midrule
A2 & \url{www.snopes.com/2015/11/23/obama-dropped-leaflets-warning-isis-airstrikes/}\\
\midrule
\multirow{10}{*}{B1} &
\url{en.mediamass.net/people/alan-rickman/deathhoax.html} \\
&\url{www.disclose.tv/forum/david-bowie-alan-rickman-death-hoax-100-staged-t108254.html}\\
&\url{worldtruth.tv/david-bowie-and-alan-rickman-death-hoax-100-staged/}\\
&\hangpara{1.5em}{1}\url{beforeitsnews.com/alternative/2016/01/alan-rickman-the-curse-of-the-69-takes-another-victim-january-man-predicts-his-death-video-3277444.html}\\
&\hangpara{1.5em}{1}\url{beforeitsnews.com/celebrities/2016/01/david-bowie-alan-rickman-death-hoax-100-staged-both-69-died-from-cancer-2474208.html}\\
&\url{age-69.beforeitsnews.com/alternative/2016/01/harry-potter-star-alan-rickman-dead-at-age-69-3277454.html}\\
&\hangpara{1.5em}{1}\url{from-cancer.beforeitsnews.com/celebrities/2016/01/david-bowie-alan-rickman-death-hoax-100-staged-both-69-died-from-cancer-2474208.html}\\
\midrule
\multirow{2}{*}{B2} &
\url{www.snopes.com/2016/01/14/alan-rickman-dies-at-69/}\\
&\url{www.snopes.com/alan-rickman-potter-meme/}\\
\bottomrule
\end{tabular}
\label{tab:urls}
\end{table*}

\begin{figure*}[t]
\centering
\includegraphics[width=.9\textwidth]{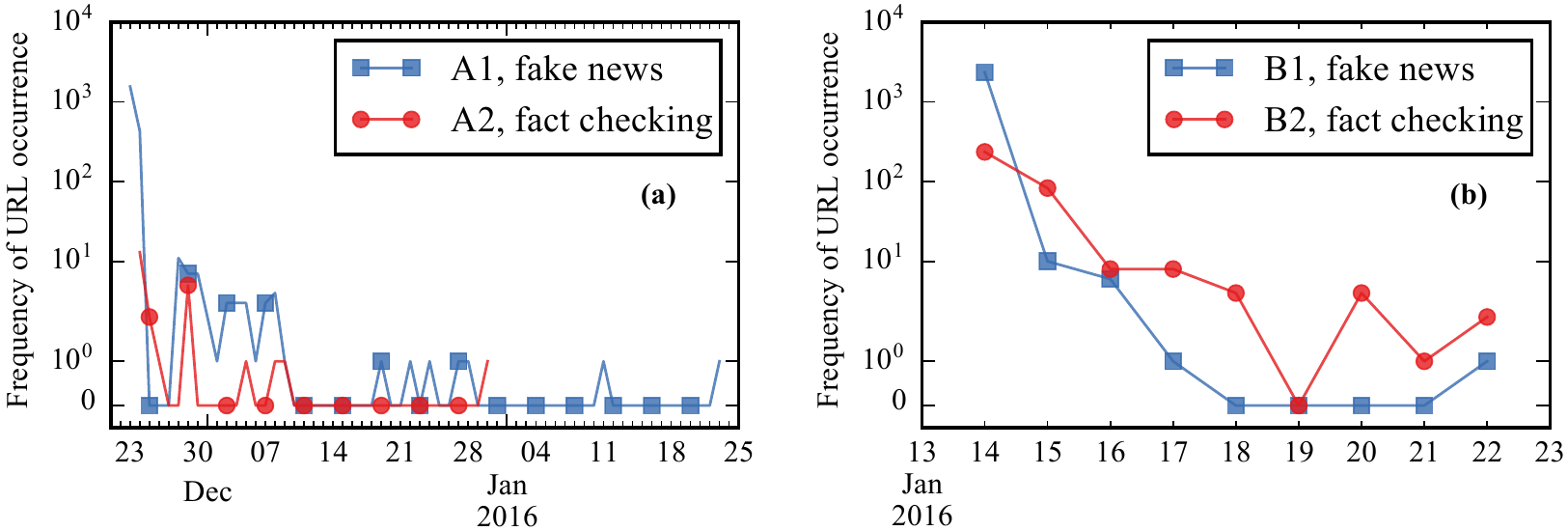}
\caption{Daily volume of tweets for (a) A1 and A2 (cf. Table~\ref{tab:urls}); (b) B1 and B2. Semi-log scale was used for the plot.}
\label{fig:examples}
\end{figure*}

Table~\ref{tab:urls} displays the two URLs (A1, A2) used in the first strategy. The reported story focuses on the current Syrian conflict, and contains several inaccuracies that were debunked in a piece by Snopes.com. For the second strategy, we focused on a recent event: the death of famous actor Alan Rickman on January 14, 2016. We used the keywords `alan' and `rickman' to match URLs from our database, and found 15 matches (B1) among fake news sources and two from fact-checking ones (B2). The fake news, in particular, were spreading the rumor that the actor had \emph{not} died. Fig.~\ref{fig:examples} plots the volume of tweets containing URLs from both strategies. Despite low data volumes, the spikes of activity and the successive decay show fairly strong alignment. 

\subsection{User Activity and URL Popularity}

\begin{figure*}
\centering
\includegraphics[width=\textwidth]{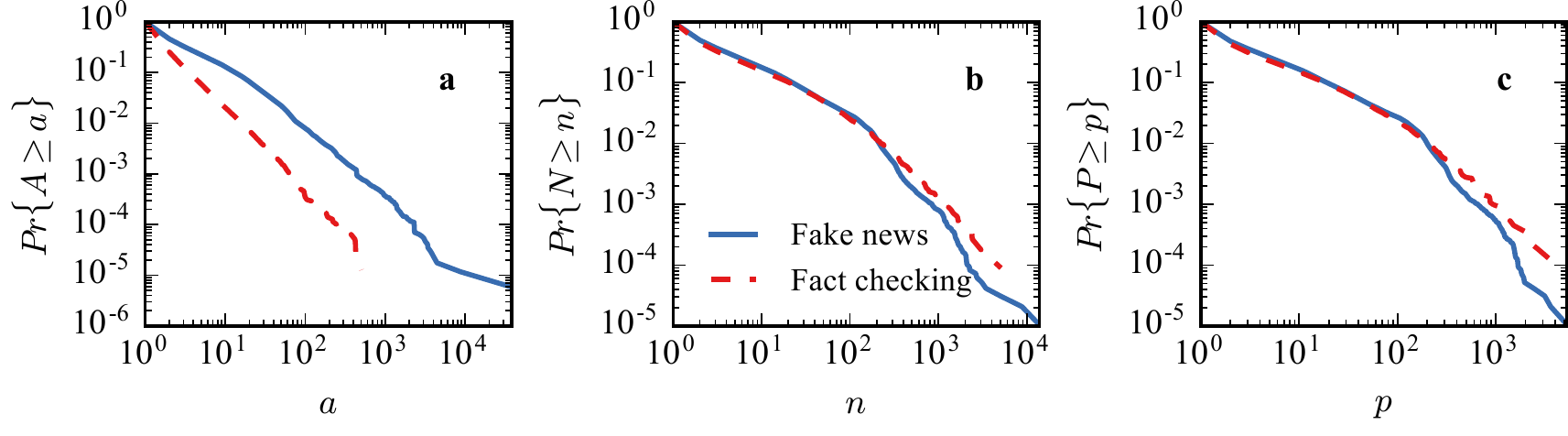}
\caption{Complementary cumulative distribution function (CCDF)  of (\textbf{a}) user activity $a$ (tweets per user)  (\textbf{b}) URL popularity $n$ (tweets per URL) and (\textbf{c}) URL popularity $p$ (users per URL).}
\label{fig:three_ccdfs}
\end{figure*}

We measure the activity $a$ of users by counting the number of tweets they posted, and the popularity of a given story (either fake news or fact checking) by counting either the total number $n$ of times its URL was tweeted, or the total number $p$ of people who tweeted it.  Fig.~\ref{fig:three_ccdfs} shows that these quantities display heavy-tailed, power-law distributions $P(x) \sim x^{-\gamma}$. 
We estimated the power-law decay exponents,  obtaining the following results: for user activity $\gamma_a^{\rm fn}=2.3$, $\gamma_a^{\rm fc}=2.7$; for URL popularity by tweets (tail fit for $n\geq 200$) $\gamma_n^{\rm fn}=2.7$, $\gamma_n^{\rm fc}=2.5$; and for URL popularity by users (tail fit for $p\geq 200$) $\gamma_p^{\rm fn}=2.9$, $\gamma_p^{\rm fc}=2.5$. These observations suggest that fake news and fact checking have similar popularity profiles, with fake news being spread by accounts that, in some cases, can generate huge numbers of tweets.
While it is expected that active users are responsible for producing a majority of news shares, the strong difference between fake news and fact checking deserves further scrutiny.

In Twitter there are four types of content: \emph{original} tweets, \emph{retweets}, \emph{quotes}, and \emph{replies}. In our data, original tweets and retweets were the most common category (80--90\%) while quotes and replies correspond to only 10--20\% of the total, usually with slightly more replies than quotes. However, we observe differences between fact checking and misinformation tweets. The first is that there are more replies and quotes among fact checking tweets ($>20\%$) than misinformation ($\approx 10\%$), suggesting that fact checking is a more \emph{conversational} task. 

\begin{figure}[h!]
\centering
\includegraphics[width=\columnwidth]{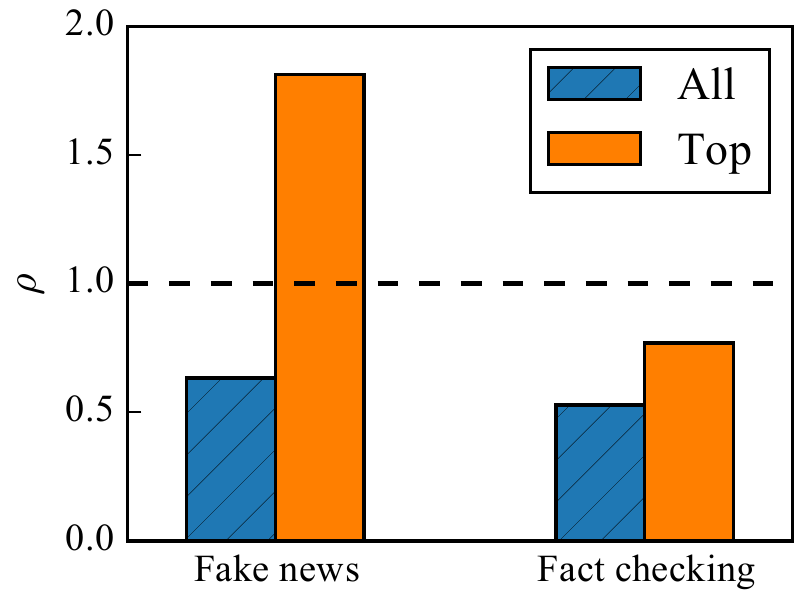}
\caption{Ratio of original tweets to retweets for all vs. top active users.}
\label{fig:type_dist}
\end{figure}

The second difference has to do with how content generation is shared among the top active users and the remaining user base. To investigate this difference, we select for both fake news and fact checking the tweets generated by the top active users, which we define as the 1\% most active user by number of tweets. 
%(i.e., the 99th percentile of users in Fig.~\ref{fig:three_ccdfs}(a)). 
%
In Fig.~\ref{fig:type_dist} we plot the ratio $\rho$ between original tweets and retweets for all users and top active ones. For all users, the ratio is similar; there are more retweets than original tweets. This is also the case for the top spreaders of fact checking. However, for top spreaders of fake news, this ratio is much higher: these users do not retweet as much but post many original messages promoting the misinformation.

Taken together, these observations strongly suggest that rumor-mongering is dominated by few very active accounts that bear the brunt of the promotion and spreading of misinformation, whereas the propagation of fact checking is a more distributed, grass-roots activity. 

\section{Conclusions \& Future Work}

Social media provide excellent examples of marketplaces of attention where different memes vie for the limited time of users~\cite{Ciampaglia2015a}. A scientific understanding of the dynamics of the Web is increasingly critical~\cite{Berners-Lee2006}, and the dynamics of online news consumption exemplify this need, as the risk of massive uncontrolled misinformation grows. Our upcoming Hoaxy platform for the automatic tracking of online misinformation may provide an important tool for the study of these phenomena. Our preliminary results suggest an interesting interplay between fake news promoted by few very active accounts, and grass-roots responses that spread fact checking information several hours later. 

In the future we plan to study the active spreaders of fake news to see if they are likely social bots~\cite{socialbots, subrahmanian2016darpa}. We will also expand our analysis to a larger set of news stories and investigate how the lag between misinformation and fact checks varies for different types of news. 
%Currently, our system allows retrospective analysis of rumor spreading on social media. Building on the present work, we hope that systems of this kind will be able to perform automatic tracking of false news.

% \newpage
\subsection*{Acknowledgments}

CS was supported by the China Scholarship Council while visiting the Center for Complex Networks and Systems Research at the Indiana University School of Informatics and Computing. GLC acknowledges support from the Indiana University Network Science Institute (\url{iuni.iu.edu}) and from the Swiss National Science Foundation (PBTIP2--142353). This work was supported in part by the NSF (award CCF-1101743) and the J.S. McDonnell Foundation.

\bibliographystyle{abbrv}
\bibliography{references}

\end{document}